# Hierarchical Bayesian Modeling for Uncertainty Quantification and Reliability Updating using Data

Xinyu Jia[1], Weinan Hou[1], Costas Papadimitriou[2, *]

[1] *State Key Laboratory of Reliability and Intelligence of Electrical Equipment, School of Mechanical Engineering, Hebei University of Technology, 300401, Tianjin, China*
[2] *Department of Mechanical Engineering, University of Thessaly, 38334 Volos, Greece*

**ABSTRACT**
Quantifying uncertainty and updating reliability are essential for ensuring the safety and performance of engineering systems. This study develops a hierarchical Bayesian modeling (HBM) framework to quantify uncertainty and update reliability using data. By leveraging the probabilistic structure of HBM, the approach provides a robust solution for integrating model uncertainties and parameter variability into reliability assessments. The framework is applied to a linear mathematical model and a dynamical structural model. For the linear model, analytical solutions are derived for the hyper parameters and reliability, offering an efficient and precise means of uncertainty quantification and reliability evaluation. In the dynamical structural model, the posterior distributions of hyper parameters obtained from the HBM are used directly to update the reliability. This approach relies on the updated posteriors to reflect the influence of system uncertainties and dynamic behavior in the reliability predictions. The proposed approach demonstrates significant advantages over traditional Bayesian inference by addressing multi-source uncertainty in both static and dynamic contexts. This work highlights the versatility and computational efficiency of the HBM framework, establishing it as a powerful tool for uncertainty quantification and reliability updating in structural health monitoring and other engineering applications.

## 1. Introduction

Uncertainty quantification and reliability updating play an essential role in the design, operation, and maintenance of engineering systems [1]. In real-world applications, uncertainties arise from various sources, including material properties, environmental and operational conditions, manufacturing variability, and modeling errors. These uncertainties can significantly impact the performance, safety, and longevity of engineering systems, making it crucial to quantify and account for such uncertainties [2]. Reliability updating, in particular, involves leveraging new data to refine reliability predictions, ensuring they reflect the current state of the system [3]. This is particularly important for systems subjected to dynamic conditions, where operational demands and environmental factors may evolve over time. Accurate uncertainty quantification and reliability updating not only enhance decision-making but also improve the cost-efficiency of maintenance strategies and reduce the risk of failures in critical systems.

Bayesian inference has become a popular approach for addressing uncertainty in engineering analysis due to its strong theoretical foundation and flexibility [4–6]. It allows for the integration of prior knowledge-derived from historical data, expert judgment, or physical models-with newly observed data to infer parameters of interest. This probabilistic approach provides not just point estimates but also distributions that characterize uncertainty, enabling us to make informed decisions under uncertainty. By incorporating all sources of evidence in a mathematically consistent manner, Bayesian inference has been successfully applied across various domains, including structural health monitoring [7–9], prognostics [10–12], and risk assessment [13,14]. In the context of reliability analysis, Bayesian methods naturally facilitate reliability updating as new information becomes available. Different strategies have been developed for reliability analysis. For example, the asymptotic approximation based on Laplace approximation [15], and the first- and second-order reliability methods [16] have been employed for reliability calculating. These approaches often provide computational efficiency when data is sufficient for reliability updating. In addition, the sampling approaches, which have been extensively explored in recent decades, have shown their advantages due to high accuracy in terms of the reliability calculations. Some relevant contributions on the sampling algorithms could be standard Monte Carlo algorithms [17,18] and advanced sampling techniques including importance sampling [19–23], BUS algorithm (Bayesian updating with structural reliability methods) [24], subset simulations [25,26] with Gibbs sampling [27], and other advanced sampling approaches [28–32]. Besides, the surrogate modeling, mainly Gaussian process, has been also explored for the context of reliability updating within the Bayesian perspective [33,34].

Despite its advantages, traditional Bayesian inference methods are not without limitations. A common issue arises from the underestimation of parameter uncertainties, especially when data appears in varying conditions or when overly simplistic prior distributions are used [35,36]. For example, the datasets could be collected within different environmental conditions, or due to manufacturing or assembling process, or varied due to modeling error [37]. The estimates of the reliability using these datasets could vary because the uncertainty due to variability exists. The classical Bayesian approach may result in unreliable results if one combines the datasets and neglects such a variability. This is not due to the use of different sampling approaches but due to the fact that the parameter uncertainty would be decreased as the number of data increases [38]. This underestimation can result in overconfident predictions, leading to reliability estimates that fail to account for the true variability and risks present in the system. Additionally, traditional Bayesian approaches sometimes rely on computational approximations, such as Laplace or variational approximations, which may introduce biases and further exacerbate the underestimation of uncertainty. These limitations highlight the need for more robust approaches that can better capture the complex and multi-source uncertainties inherent in engineering systems.

Hierarchical Bayesian modeling (HBM) provides a powerful extension to the traditional Bayesian framework, addressing the shortcomings of the traditional Bayesian framework [39]. The HBM framework has been developed in several scientific fields, such as structural dynamics [40–44], molecular dynamics [45–47], geotechnical engineering [48–50] to capture the uncertainty due to variability over the datasets. By introducing hierarchical structures, HBM models uncertainty at multiple levels, allowing for a more nuanced representation of variability arising from different sources, such as modeling errors, environmental influences, and manufacturing processes. This hierarchical structure enables HBM to explicitly distinguish between uncertainties associated with model parameters and those related to system variability, resulting in more realistic and reliable predictions. The flexibility of HBM supports its application to both linear and nonlinear systems, as well as static and dynamic problems. In reliability analysis, the ability of HBM to incorporate diverse datasets, such as operational data and experimental observations, makes it particularly effective for updating reliability predictions over time.

This paper advances the state-of-the-art in uncertainty quantification and reliability updating by further developing the HBM framework. The study considers two distinct modeling scenarios: (1) a linear mathematical model and (2) a dynamical structural model. For the linear case, an analytical solution is derived for the hyper parameters and reliability index, showcasing the computational efficiency. In contrast, the dynamical structural model presents challenges due to its complexity and the lack of an analytical solution. To address this, a two-stage sampling approach is employed, ensuring the accuracy of the results while maintaining computational feasibility. This work not only demonstrates the versatility of the HBM framework in handling different types of models but also highlights its ability to capture and quantify uncertainties comprehensively.

The structure of the paper is organized as follows. Section 2 provides an overview of the HBM framework, including its mathematical formulation and fundamental principles. Section 3 focuses on the linear mathematical model presenting the derivation of an analytical solution for uncertainty quantification and reliability, as well as a dynamical structural model describing the challenges associated with its complexity and the use of a two-stage sampling procedure to ensure accurate results. Section 4 presents two numerical examples to validate the proposed framework, demonstrating its effectiveness across both linear and dynamical cases. Section 5 concludes the paper by summarizing the key findings, discussing their implications, and suggesting potential directions for future research.

## 2. Brief review of the HBM framework

Hierarchical Bayesian modeling (HBM) is a statistical approach that can capture the uncertainty due to variability arising from modeling errors, environmental and operational conditions, as well as manufacturing and assembling processes. It is particularly useful for systems where uncertainty is inherent due to these sources of variability, and effective for dealing with multiple datasets and incorporating prior knowledge. The core of HBM lies in defining prior distributions for both model parameters and hyper parameters, followed by updating these priors with observed datasets to obtain posterior distributions. Consider a general model where the parameters $\boldsymbol{\theta} \in \mathbb{R}^{N_\theta}$ are of interest, where $N_\theta$ is the total number of unknown parameters in the set $\boldsymbol{\theta}$. In HBM, the model parameters typically follow a Gaussian prior distribution [51]:

$$p(\boldsymbol{\theta}|\boldsymbol{\psi}) = N(\boldsymbol{\theta}|\boldsymbol{\mu}_\theta, \boldsymbol{\Sigma}_\theta) \tag{1}$$

with hyper parameter $\boldsymbol{\psi}$ consisting of the hyper mean $\boldsymbol{\mu}_\theta \in \mathbb{R}^{N_\theta}$ and hyper covariance matrix $\boldsymbol{\Sigma}_\theta \in \mathbb{R}^{N_\theta \times N_\theta}$ to be identified from the available multiple datasets $\mathbf{D} = \{\mathbf{D}_i; i=1,2,\ldots,N_D\}$. Let $\boldsymbol{\theta}_i$ denote an independent realization of the parameter $\boldsymbol{\theta}$ from Gaussian distribution $N(\boldsymbol{\mu}_\theta, \boldsymbol{\Sigma}_\theta)$, corresponding to the $i$-th dataset $\mathbf{D}_i$. The joint distribution of the overall parameters $p(\boldsymbol{\theta}, \boldsymbol{\psi}|\mathbf{D})$ can be expressed [52] using the Bayes' theorem as:

$$p(\boldsymbol{\theta}, \boldsymbol{\psi}|\mathbf{D}) \propto p(\mathbf{D}|\boldsymbol{\theta}, \boldsymbol{\psi}) p(\boldsymbol{\theta}|\boldsymbol{\psi}) p(\boldsymbol{\psi}) \tag{2}$$

where $p(\boldsymbol{\psi})$ is the prior distribution of the hyper parameter, and $p(\boldsymbol{\theta}|\boldsymbol{\psi})$ and $p(\mathbf{D}|\boldsymbol{\theta}, \boldsymbol{\psi})$ is the joint Gaussian prior distribution of the model parameters and the joint likelihood function. Given the independence between datasets, Eq. (2) is simplified as:

$$p(\boldsymbol{\theta}, \boldsymbol{\psi}|\mathbf{D}) \propto p(\boldsymbol{\psi}) \prod_{i=1}^{N_D} p(\boldsymbol{\theta}_i|\boldsymbol{\psi}) p(\mathbf{D}_i|\boldsymbol{\theta}_i) \tag{3}$$

where $p(\mathbf{D}_i|\boldsymbol{\theta}_i)$ is the likelihood function that corresponds to the $i$-th dataset $\mathbf{D}_i$.

Subsequently, the posterior distribution of the hyper parameter $p(\boldsymbol{\psi}|\mathbf{D})$ can be obtained by marginalizing the distribution $p(\boldsymbol{\theta}, \boldsymbol{\psi}|\mathbf{D})$ over the model parameters $\boldsymbol{\theta}$. Two cases, including a linear mathematical model and a dynamical structure, will be discussed in the next section. In the first linear case, the property of the linear function will be utilized to derive an analytical solution for the posterior distribution of the model parameters, which corresponds to each dataset. This solution will be used to update the reliability of the system by incorporating the latest data. The analytical approach simplifies the process, offering an efficient means for reliability analysis, as the posterior distribution can be directly computed without the need for complex sampling methods. In the second case, a dynamical structure is considered, where the model exhibits more complexity due to the time-varying behavior and interdependencies among system parameters. Here, a sampling procedure will be employed to estimate the posterior distributions of the hyper parameters. This approach accommodates the non-linearity and uncertainty inherent in the dynamical system, ensuring that the reliability updates reflect both the variability of the system and the uncertainty from the load excitation. Through this combination of analytical and sampling methods, the proposed framework provides a versatile approach to uncertainty quantification and reliability updating. Details are given in the next section.

# 3. Uncertainty quantification and reliability updating using HBM
## 3.1. A linear mathematical model

The linear model considered in this study is expressed as:

$$g(\boldsymbol{\theta}) = \mathbf{A}^T \boldsymbol{\theta} \tag{4}$$

where $\mathbf{A}$ is a known system matrix, $\boldsymbol{\theta}$ is the model parameter vector, and the observed data $\mathbf{y}$ follows the likelihood function:

$$\mathbf{y} \sim N(\mathbf{A}^T \boldsymbol{\theta}, \sigma^2 \mathbf{I}) \tag{5}$$

where $\sigma^2$ is the variance of the measurement noise. Expanding the likelihood function leads to the following form:

$$p(\mathbf{y} \mid \boldsymbol{\theta}) \propto \exp\left(-\frac{1}{2\sigma^2}\left[\mathbf{y} - \mathbf{A}^T\boldsymbol{\theta}\right]^T \left[\mathbf{y} - \mathbf{A}^T\boldsymbol{\theta}\right]\right) \tag{6}$$

Eq. (6) can be also seen as a function of the model parameters $\boldsymbol{\theta}$, and can thus be written in a Gaussian form:

$$p(\mathbf{y} \mid \boldsymbol{\theta}) \propto N(\boldsymbol{\theta} \mid \boldsymbol{\theta}^*, \boldsymbol{\Sigma}_{\boldsymbol{\theta}}^*) \tag{7}$$

where the mean $\boldsymbol{\theta}^*$ and the covariance matrix $\boldsymbol{\Sigma}_{\boldsymbol{\theta}}^*$ are given by:

$$\begin{aligned}\boldsymbol{\theta}^* &= \boldsymbol{\Sigma}_{\boldsymbol{\theta}}^* \mathbf{A} \sigma^{-2} \mathbf{y} \\ \boldsymbol{\Sigma}_{\boldsymbol{\theta}}^* &= \sigma^2 \left(\mathbf{A}\mathbf{A}^T\right)^{-1}\end{aligned} \tag{8}$$

Assume that the datasets $\mathbf{D}$ are taken from the linear model with the $i$-th dataset $\mathbf{D}_i$ equals to $\mathbf{y}_i$, the likelihood function $p(\mathbf{D}_i \mid \boldsymbol{\theta}_i)$ is then rewritten as:

$$p(\mathbf{D}_i \mid \boldsymbol{\theta}_i) \propto N(\boldsymbol{\theta}_i \mid \boldsymbol{\theta}_i^*, \boldsymbol{\Sigma}_{\boldsymbol{\theta}_i}^*) \tag{9}$$

where the mean $\boldsymbol{\theta}_i^*$ and the covariance matrix $\boldsymbol{\Sigma}_{\boldsymbol{\theta}_i}^*$ for each dataset are calculated using the solutions in Eq. (8) by replacing by $\boldsymbol{\theta}^*$, $\boldsymbol{\Sigma}_{\boldsymbol{\theta}}^*$ and $\mathbf{y}$ by $\boldsymbol{\theta}_i^*$, $\boldsymbol{\Sigma}_{\boldsymbol{\theta}_i}^*$ and $\mathbf{y}_i$, respectively. It is notable that the variance of the measurement noise is assumed to be the same for each dataset. Eq. (3) can then be rewritten as:

$$p(\boldsymbol{\theta}, \boldsymbol{\mu}_{\boldsymbol{\theta}}, \boldsymbol{\Sigma}_{\boldsymbol{\theta}} \mid \mathbf{D}) \propto p(\boldsymbol{\mu}_{\boldsymbol{\theta}}, \boldsymbol{\Sigma}_{\boldsymbol{\theta}}) \prod_{i=1}^{N_D} N(\boldsymbol{\theta}_i \mid \boldsymbol{\mu}_{\boldsymbol{\theta}}, \boldsymbol{\Sigma}_{\boldsymbol{\theta}}) N(\boldsymbol{\theta}_i \mid \boldsymbol{\theta}_i^*, \boldsymbol{\Sigma}_{\boldsymbol{\theta}_i}^*) \tag{10}$$

Subsequently, the posterior distributions can be solved as:

$$p(\boldsymbol{\mu}_{\boldsymbol{\theta}}, \boldsymbol{\Sigma}_{\boldsymbol{\theta}} \mid \mathbf{D}) \propto p(\boldsymbol{\mu}_{\boldsymbol{\theta}}, \boldsymbol{\Sigma}_{\boldsymbol{\theta}}) \prod_{i=1}^{N_D} \int N(\boldsymbol{\theta}_i \mid \boldsymbol{\mu}_{\boldsymbol{\theta}}, \boldsymbol{\Sigma}_{\boldsymbol{\theta}}) N(\boldsymbol{\theta}_i \mid \boldsymbol{\theta}_i^*, \boldsymbol{\Sigma}_{\boldsymbol{\theta}_i}^*) d\boldsymbol{\theta}_i \tag{11}$$

The integral in Eq. (11) can be solved by using the rule given in [36]:

$$\int N(\boldsymbol{\theta}_i \mid \boldsymbol{\mu}_{\boldsymbol{\theta}}, \boldsymbol{\Sigma}_{\boldsymbol{\theta}}) N(\boldsymbol{\theta}_i \mid \boldsymbol{\theta}_i^*, \boldsymbol{\Sigma}_{\boldsymbol{\theta}_i}^*) d\boldsymbol{\theta}_i = N(\boldsymbol{\mu}_{\boldsymbol{\theta}} \mid \boldsymbol{\theta}_i^*, \boldsymbol{\Sigma}_{\boldsymbol{\theta}} + \boldsymbol{\Sigma}_{\boldsymbol{\theta}_i}^*) \tag{12}$$

Therefore, the posterior distributions of the hyper parameters are computed in the following form:

$$p(\boldsymbol{\mu}_{\boldsymbol{\theta}}, \boldsymbol{\Sigma}_{\boldsymbol{\theta}} \mid \mathbf{D}) \propto p(\boldsymbol{\mu}_{\boldsymbol{\theta}}, \boldsymbol{\Sigma}_{\boldsymbol{\theta}}) \prod_{i=1}^{N_D} N(\boldsymbol{\mu}_{\boldsymbol{\theta}} \mid \boldsymbol{\theta}_i^*, \boldsymbol{\Sigma}_{\boldsymbol{\theta}} + \boldsymbol{\Sigma}_{\boldsymbol{\theta}_i}^*) \tag{13}$$

It is noted that analytical solution for the posterior distributions of the hyper parameters is achieved as shown in Eq. (13). It provides computational efficiency, as it eliminates the need for time-consuming methods like Markov Chain Monte Carlo (MCMC) or any other samplings. Also, it is beneficial for reliability updating as it allows for fast updates to reliability assessments without the need for time-intensive sampling.

To compute the reliability index, a scalar performance function is typically required. One common way is to define a scalar limit state function based on a critical component or direction in the parameter space:

$$G(\boldsymbol{\theta}) = b - \mathbf{c}^T \mathbf{A}^T \boldsymbol{\theta} = b - (\mathbf{A}\mathbf{c})^T \boldsymbol{\theta} \tag{14}$$

where b is the maximum exceedance level, **c** is a weighting vector (or a direction vector) and **Ac** acts as a transformed system matrix. The failure probability conditional on the samples $\boldsymbol{\mu}_\theta^{(m)}$ and $\boldsymbol{\Sigma}_\theta^{(m)}$ is given by:

$$F^{(m)}(\boldsymbol{\mu}_\theta^{(m)}, \boldsymbol{\Sigma}_\theta^{(m)}) = \int_{G(\boldsymbol{\theta}) \leq 0} N(\boldsymbol{\theta} \mid \boldsymbol{\mu}_\theta^{(m)}, \boldsymbol{\Sigma}_\theta^{(m)}) d\boldsymbol{\theta} = \Phi(-\beta^{(m)}) \tag{15}$$

where $\beta^{(m)}$ is the reliability index condition on a hyper parameter sample

$$\beta^{(m)} = \frac{(b - (\mathbf{Ac})^T)\boldsymbol{\mu}_\theta^{(m)}}{\sqrt{(\mathbf{Ac})^T \boldsymbol{\Sigma}_\theta^{(m)} (\mathbf{Ac})}} \tag{16}$$

Therefore, the posterior probability of failure given multiple datasets is

$$\Pr(F \mid \mathbf{D}) = \frac{1}{N_s} \sum_{m=1}^{N_s} \Phi(-\beta^{(m)}) \tag{17}$$

which is the average of the failure probabilities over all hyper parameter samples.

### *3.2. Structural dynamical model*

This second case applies the HBM framework to a dynamical structural model, which differs from the previous linear case. Unlike linear models, where an analytical solution for the posterior distribution is achievable, dynamic models are more complex, and an analytical solution is not feasible. While several studies have used asymptotic approximations [36,44] or variational inference [52] to obtain semi-analytical solutions for such systems, this work employs a two-stage sampling approach to ensure the accuracy of the HBM framework [53,54]. The two-stage approach helps capture the full posterior distribution of the hyper parameters, ensuring accurate uncertainty quantification and reliability updating using multiple datasets.

Consider a dynamical structural model which describes the behavior of a structural system under varying conditions, and let $\boldsymbol{\theta} \in R^{N_\theta}$ be the set of material/structural model parameters to be estimated using data, and $N_\theta$ is the total number of the unknown parameters in the set $\boldsymbol{\theta}$. The data used for this model can take the form of modal properties, such as natural frequencies, mode shapes, and damping ratios, or time histories, including displacement, velocity, acceleration or strain data over time. These data types are essential for capturing the dynamic response of the structure and are incorporated into the HBM framework for parameter estimation and uncertainty quantification.

Let the dataset be defined as $\mathbf{D} = \{\mathbf{D}_i; i = 1, 2, \ldots, N_D\}$, where $N_D$ is the number of datasets which is the same with the first case. Followed by Eq. (3), the posterior distribution of the hyper parameters $p(\boldsymbol{\psi} \mid \mathbf{D})$ can be further computed by marginalization of the distribution $p(\boldsymbol{\theta}, \boldsymbol{\psi} \mid \mathbf{D})$ over the model parameters $\boldsymbol{\theta}$:

$$p(\boldsymbol{\psi} \mid \mathbf{D}) \propto p(\boldsymbol{\psi}) \prod_{i=1}^{N_D} \int p(\boldsymbol{\theta}_i \mid \boldsymbol{\psi}) p(\mathbf{D}_i \mid \boldsymbol{\theta}_i) d\boldsymbol{\theta}_i \tag{18}$$

The integral in Eq. (18) can be solved using MC simulation, given by:

$$\int p(\boldsymbol{\theta}_i \mid \boldsymbol{\psi}) p(\mathbf{D}_i \mid \boldsymbol{\theta}_i) d\boldsymbol{\theta}_i \approx \frac{1}{N_s} \sum_{l=1}^{N_s} p(\boldsymbol{\theta}_i^{(l)} \mid \boldsymbol{\psi}) \tag{19}$$

where $\boldsymbol{\theta}_i^{(l)}$ is the *l*-th sample taken from the likelihood function $p(\mathbf{D}_i \mid \boldsymbol{\theta}_i)$ for the *i*-th dataset, which can be obtained using the Transitional MCMC sampling algorithm [28,54]. $N_s$ is the number of samples for the model parameters. Substituting Eq. (19) into Eq. (18) leads to the following form:

$$p(\boldsymbol{\psi} \mid \mathbf{D}) \propto p(\boldsymbol{\psi}) \prod_{i=1}^{N_D} \frac{1}{N_s} \sum_{l=1}^{N_s} p(\boldsymbol{\theta}_i^{(l)} \mid \boldsymbol{\psi}) \tag{20}$$

By replacing the hyper parameter $\boldsymbol{\psi}$ with hyper mean and hyper covariance matrix, Eq. (20) takes the

form:
$$p(\boldsymbol{\mu_\theta}, \boldsymbol{\Sigma_\theta} | \mathbf{D}) \propto p(\boldsymbol{\mu_\theta}, \boldsymbol{\Sigma_\theta}) \prod_{i=1}^{N_D} \frac{1}{N_s} \sum_{l=1}^{N_s} N(\boldsymbol{\theta}_i^{(l)} | \boldsymbol{\mu_\theta}, \boldsymbol{\Sigma_\theta}) \quad (21)$$

Eq. (21) can be then solved by a two-stage sampling approach. In the first stage, the sample $\boldsymbol{\theta}_i^{(l)}$ needs to be generated from the distribution $p(\mathbf{D}_i | \boldsymbol{\theta}_i)$. It thus requires the model runs which is the most expensive computational part. In the second step, it does not require the model runs and only the samples from the first step need to be incorporated into Eq. (21). The samples in the second step are also generated using the Transitional MCMC sampling algorithm [28,54]. Once the samples of the hyper parameters are available, the reliability could be conducted conditional on these samples.

In this study, two cases are considered for reliability updating where the first case only accounts for the means of the hyper parameters while the second case considers all the samples of the hyper parameters. The updates of the reliability between two cases are expected to be close when given sufficient number of datasets. This is because the uncertainty of the hyper parameters would decrease as the number of datasets increases [52]. Given a large number of datasets, the posterior distribution would be a peaked distribution for which their uncertainty does not play a role for reliability updating. This scenario can be further demonstrated by the insights given in Ref. [52] in the sense of updating reliability.

Considering only the means of the hyper parameters, the posterior probability of failure given multiple datasets is expressed as:

$$\Pr(F|\mathbf{D}) = \int_{\substack{\boldsymbol{\theta} \in \mathbb{R}^{N_\theta} \\ \boldsymbol{\varphi} \in \mathbb{R}^{N_\varphi}}} p(\boldsymbol{\varphi}) \mathrm{I}_F(\boldsymbol{\theta}, \boldsymbol{\varphi}) p(\boldsymbol{\theta} | \mathbf{D}) \mathrm{d}\boldsymbol{\varphi} \mathrm{d}\boldsymbol{\theta} = \int_{\substack{\boldsymbol{\theta} \in \mathbb{R}^{N_\theta} \\ \boldsymbol{\varphi} \in \mathbb{R}^{N_\varphi}}} p(\boldsymbol{\varphi}) \mathrm{I}_F(\boldsymbol{\theta}, \boldsymbol{\varphi}) N(\boldsymbol{\theta} | \bar{\boldsymbol{\mu}}_\theta, \bar{\boldsymbol{\Sigma}}_\theta) \mathrm{d}\boldsymbol{\varphi} \mathrm{d}\boldsymbol{\theta} \quad (22)$$

where $\bar{\boldsymbol{\mu}}_\theta$ and $\bar{\boldsymbol{\Sigma}}_\theta$ are the means of the hyper parameters, $\mathrm{I}_F(\boldsymbol{\theta}, \boldsymbol{\varphi})$ is the indicator function such that $\mathrm{I}_F(\boldsymbol{\theta}, \boldsymbol{\varphi}) = 1$ when the limit state function satisfies $G(\boldsymbol{\theta}, \boldsymbol{\varphi}) \leq 0$, otherwise it will be zero. Note that $\boldsymbol{\varphi}$ is the set of parameters associated with other uncertainties (e.g. input related uncertainties) that are not included in the parameter set $\boldsymbol{\theta}$ to be inferred by the HBM framework. It is assumed that $\boldsymbol{\varphi}$ is uncorrelated to the model parameters and a PDF $p(\boldsymbol{\varphi})$ is assigned to quantify uncertainties in $\boldsymbol{\varphi}$. Subset simulation is used in this study to compute the probability of failure using multiple datasets.

When all the samples of the hyper parameters are considered, the posterior probability of failure given multiple datasets is given by:

$$\Pr(F|\mathbf{D}) = \int_{\substack{\boldsymbol{\theta} \in \mathbb{R}^{N_\theta} \\ \boldsymbol{\varphi} \in \mathbb{R}^{N_\varphi}}} p(\boldsymbol{\varphi}) \mathrm{I}_F(\boldsymbol{\theta}, \boldsymbol{\varphi}) p(\boldsymbol{\theta} | \mathbf{D}) \mathrm{d}\boldsymbol{\varphi} \mathrm{d}\boldsymbol{\theta} = \frac{1}{N_s} \sum_{m=1}^{N_s} \int_{\substack{\boldsymbol{\theta} \in \mathbb{R}^{N_\theta} \\ \boldsymbol{\varphi} \in \mathbb{R}^{N_\varphi}}} p(\boldsymbol{\varphi}) \mathrm{I}_F(\boldsymbol{\theta}, \boldsymbol{\varphi}) N(\boldsymbol{\theta} | \boldsymbol{\mu}_\theta^{(m)}, \boldsymbol{\Sigma}_\theta^{(m)}) \mathrm{d}\boldsymbol{\varphi} \mathrm{d}\boldsymbol{\theta}$$
$$= \frac{1}{N_s} \sum_{m=1}^{N_s} F^{(m)}(\boldsymbol{\mu}_\theta^{(m)}, \boldsymbol{\Sigma}_\theta^{(m)}) \quad (23)$$

where $F^{(m)}(\boldsymbol{\mu}_\theta^{(m)}, \boldsymbol{\Sigma}_\theta^{(m)})$ is the failure probability conditional on the samples of the hyper parameters $\boldsymbol{\mu}_\theta^{(m)}$ and $\boldsymbol{\Sigma}_\theta^{(m)}$, given by:

$$F^{(m)}(\boldsymbol{\mu}_\theta^{(m)}, \boldsymbol{\Sigma}_\theta^{(m)}) = \int_{\substack{\boldsymbol{\theta} \in \mathbb{R}^{N_\theta} \\ \boldsymbol{\varphi} \in \mathbb{R}^{N_\varphi}}} p(\boldsymbol{\varphi}) \mathrm{I}_F(\boldsymbol{\theta}, \boldsymbol{\varphi}) N(\boldsymbol{\theta} | \boldsymbol{\mu}_\theta^{(m)}, \boldsymbol{\Sigma}_\theta^{(m)}) \mathrm{d}\boldsymbol{\varphi} \mathrm{d}\boldsymbol{\theta} \quad (24)$$

in which the subset simulation could be implemented to compute $F^{(m)}(\boldsymbol{\mu}_\theta^{(m)}, \boldsymbol{\Sigma}_\theta^{(m)})$.

## 4. Illustrative examples
### 4.1. HBM framework for a linear mathematical model

The linear model of Eq. (4) is taken as an example to explore the uncertainty and reliability updating using data. The model is updated by three parameters $\boldsymbol{\theta} = [\theta_1, \theta_2, \theta_3]^T$ with nominal values equaling to one. The coefficient matrix $\mathbf{A}^T$ is generated by random numbers from one to five and has dimensions $N_d \times 3$, where $N_d$ is the number of data points in a dataset. For exploring how the number of data points would affect the parameter uncertainty in a dataset, $N_d$ is selected as 50, 100, 200 and 500, respectively. The dataset is simulated based on the Gaussian distribution of the hyper mean $\mu_\theta = [1,1,1]^T$ and hyper covariance matrix $\Sigma_\theta = \text{diag}(0.05^2, 0.05^2, 0.05^2)$. Subsequently, 2% Gaussian noise is added to the measurements. Then the mean $\theta^*$ and the covariance matrix $\Sigma_\theta^*$ are solved according to Eq. (8). **Table 1** shows the changes of standard deviation ($\sigma_i, i=1,2,3$) in the covariance matrix $\Sigma_\theta^*$ corresponding to each parameter when $N_d$ takes different values. It is found that the value of standard deviation ($\sigma_i, i=1,2,3$) decreases with the increase of data points, indicating that the epistemic uncertainty decreases with the increase of data points.

**Table 1** Standard deviation at different data points

|  | $N_d = 50$ | $N_d = 100$ | $N_d = 200$ | $N_d = 500$ |
|---|---|---|---|---|
| $\sigma_1 (10^{-4})$ | 5.3115 | 5.2618 | 5.1702 | 5.0436 |
| $\sigma_2 (10^{-4})$ | 5.4899 | 5.2377 | 5.1923 | 5.1912 |
| $\sigma_3 (10^{-4})$ | 5.3262 | 5.3062 | 5.1502 | 5.0858 |

Once the mean $\theta^*$ and the covariance matrix $\Sigma_\theta^*$ are obtained for each dataset, the distribution of the hyper parameters could be computed using Eq. (13). In this example, 5000 samples were used for analysis based on the HBM method, and the data points in a dataset are chosen as 500. **Fig. 1** depicts the hyper parameter posterior distribution of model parameters for 500 datasets, with $\mu_\theta$ and $\sigma_\theta$ representing the hyper parameters associated with the model parameters. It is worth noting that the dimension of the coefficient matrix $\mathbf{A}^T$ is $500 \times 3$, that is, the number of data points in each dataset is 500. It can be seen that the mean values of the hyper parameters are in good agreement with their nominal values identified using the 500 datasets.

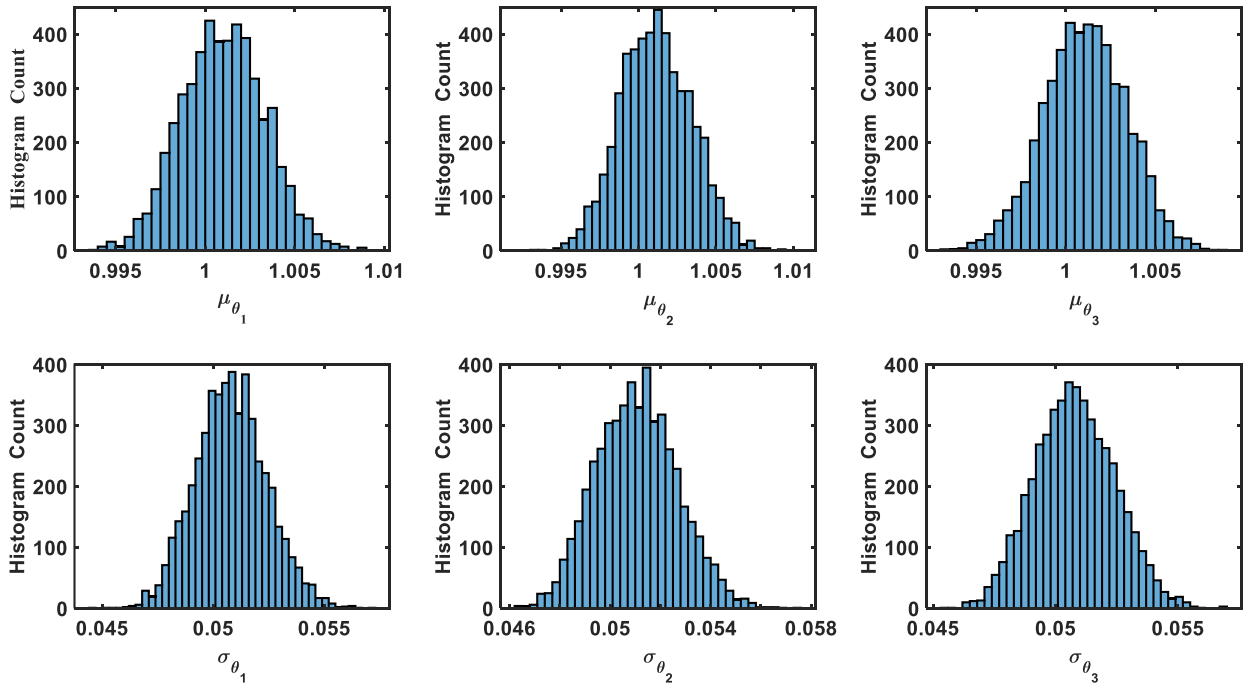

**Fig. 1** Posterior distribution of the hyper parameters using HBM method ($N_D = 500$)

In order to explore the effects of different datasets on the posterior probability distribution, different data sets are also simulated. The model is updated using four datasets of varying numbers from $N_D = 50$ to $N_D = 500$, where $N_D$ defines the number of datasets. The results in **Table 2** show the statistics of the posteriori distribution of hyper parameters obtained based on HBM method for four datasets. As can be seen from **Table 2**, the mean values of hyper mean and hyper standard deviation obtained by using different data sets based on HBM approach are close to the nominal values. It can also be seen that when given a smaller data set ($N_D = 50$), the mean of the posterior sample deviates from the nominal value of the hyper parameter. In this case, the hyper parameters have great uncertainty. However, with the increase of the number of data sets, the mean value of the hyper parameter posterior sample is closer to the nominal value of the hyper parameter, and the uncertainty of the hyper parameter decreases. The above results confirm that the uncertainty of hyper parameters decreases with the increase of the number of datasets. When given sufficient number of datasets, the mean of the hyper parameter will be closer to its nominal value. In addition, **Fig. 2** shows the PDF of hyper parameter samples of different data sets. From the results in **Fig. 2**, it can be seen that as the data set increases, the probability distribution of the hyper parameter samples takes on a narrower range of values i.e. the distribution of the corresponding samples is denser, and the uncertainty is smaller. The more data sets, the closer the hyper mean and hyper standard differences are to the nominal values. However, regardless of the size of the data set, the nominal value of the hyper parameter is included in this uncertainty. The above content is the same as the results obtained through the analysis in **Table 2**, which verifies the property that the more data sets, the smaller the uncertainty of the hyper parameters.

The posterior distribution of the model parameters can then be obtained with the samples of their hyper parameters. The results of the model parameters using the proposed HBM framework are compared with those from the classical Bayesian modeling (CBM) framework. For CBM, all the datasets are incorporated in a dataset, and 5000 samples are generated using the TMCMC algorithm for the posterior distribution of the model parameters. The results of the posterior distribution of the model parameters are given in **Fig. 3**. Good agreements are observed for the mean values of the model parameters between two approaches, while the CBM method provides an extremely thin uncertainty bound which cannot capture the uncertainty due to variability from dataset to dataset. The reason for this result is that the CBM method seriously underestimates the uncertainty of the parameters, so the probability density curve of the model parameters obtained by CBM method is narrow.

**Table 2** Statistics of hyper parameters based on HBM method

|  |  | $\theta_1$ | | $\theta_2$ | | $\theta_3$ | |
|---|---|---|---|---|---|---|---|
|  |  | $\mu_{\theta_1}$ | $\sigma_{\theta_1}$ | $\mu_{\theta_2}$ | $\sigma_{\theta_2}$ | $\mu_{\theta_3}$ | $\sigma_{\theta_3}$ |
| $N_D = 50$ | Mean value | 0.9975 | 0.0565 | 0.9969. | 0.0557 | 0.9977 | 0.0569 |
|  | Standard deviation | 0.0076 | 0.0057 | 0.0076 | 0.0057 | 0.0079 | 0.0061 |
| $N_D = 100$ | Mean value | 1.0031 | 0.0489 | 1.0034 | 0.0489 | 1.0034 | 0.0486 |
|  | Standard deviation | 0.0048 | 0.0034 | 0.0050 | 0.0034 | 0.0048 | 0.0036 |
| $N_D = 200$ | Mean value | 1.0023 | 0.0516 | 1.0025 | 0.0518 | 1.0027 | 0.0516 |
|  | Standard deviation | 0.0035 | 0.0025 | 0.0036 | 0.0025 | 0.0037 | 0.0025 |
| $N_D = 500$ | Mean value | 1.0010 | 0.0508 | 1.0011 | 0.0511 | 1.0010 | 0.0508 |
|  | Standard deviation | 0.0023 | 0.0016 | 0.0023 | 0.0016 | 0.0023 | 0.0016 |

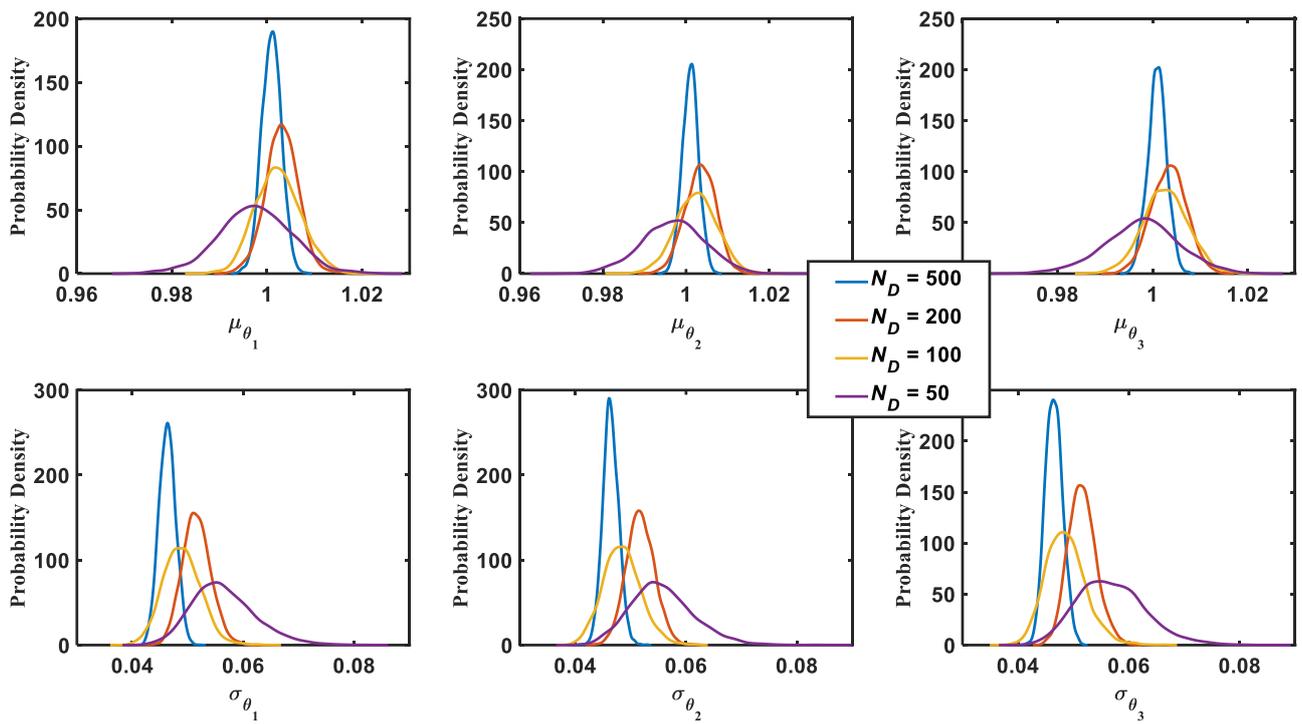

**Fig. 2** Distributions of hyper parameters corresponding to different datasets

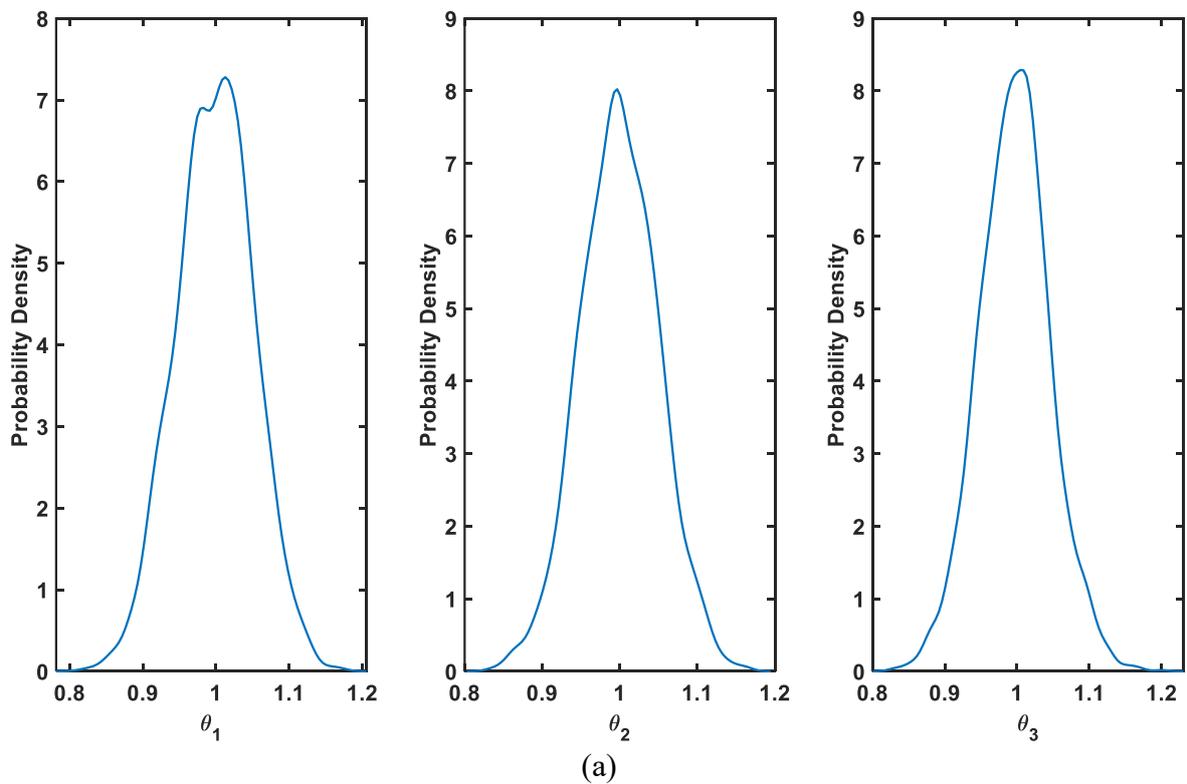

(a)

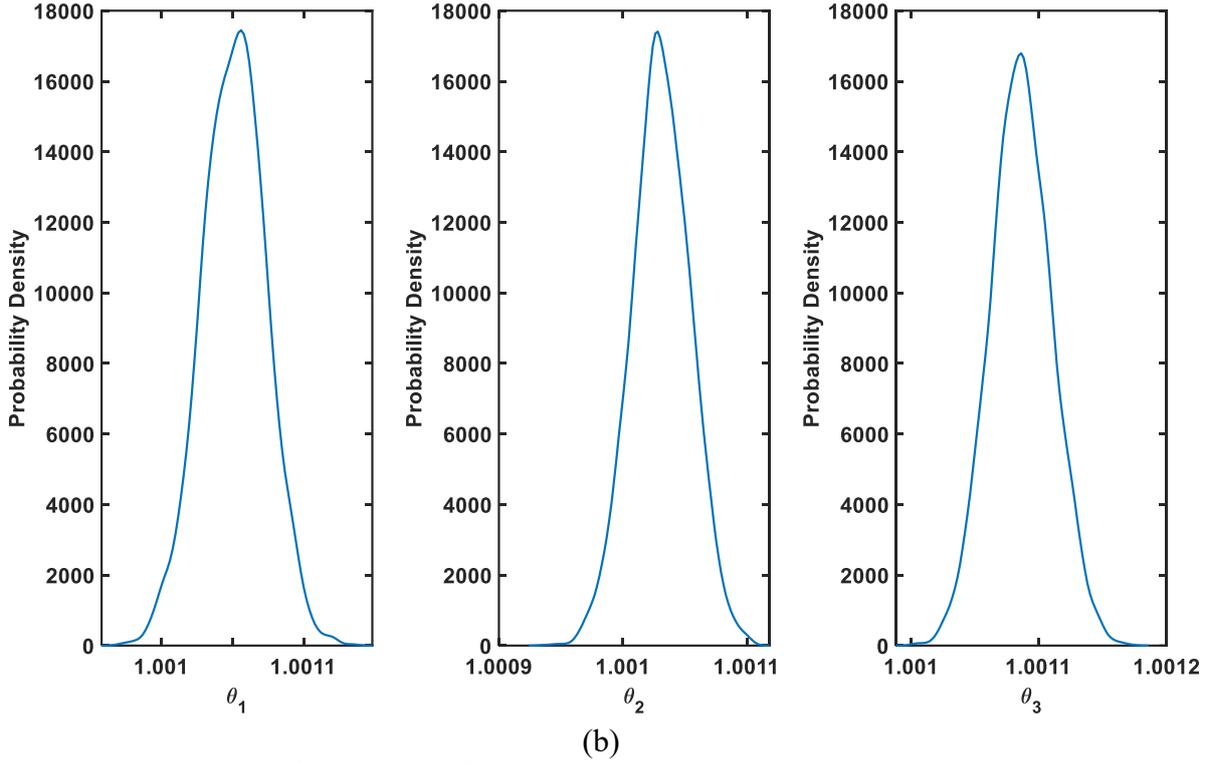
(b)

**Fig. 3** Model parameter distribution using (a) HBM method (b) CBM method

Reliability analysis is then conducted using the hyper samples given in **Fig. 1.** The limit state function is given by $G(\boldsymbol{\theta}) = b - (\mathbf{Ac})^T \boldsymbol{\theta}$ with $\mathbf{c}^T$ equal to all-one matrix with dimension $1 \times 500$. **Fig. 4** shows the failure probability based on the exceedance level b using all samples of hyper parameters obtained based on HBM method and the samples of the model parameters using CBM method. The failure probability given by the two methods differs greatly, and the values of failure probability differ by many orders of magnitude under the same vertical coordinate. The above situation is caused by different uncertainties, that is, different uncertainties lead to different reliability between two approaches. The large difference between the two in the figure is mainly because the CBM method seriously underestimates the uncertainty.

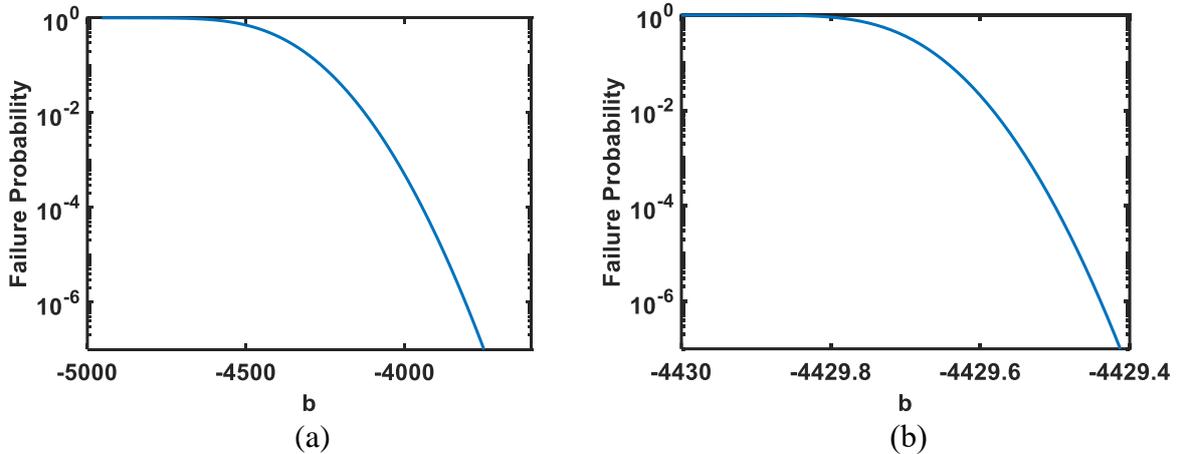
(a)          (b)

**Fig. 4** Reliability results using (a) HBM (b) CBM

### 4.2. Dynamical system: 3 degrees-of-freedom (DOF) spring mass model

Taking the three DOF spring mass model in structural dynamics as an example, the performance of the proposed method is evaluated. The structure is shown in **Fig. 5**, where the system is fixed to the left. The nominal mass ($m_0^i$) is 1 kg and the stiffness ($k_0^i$) per chain is 1800 N/m. The damping ratio of each mode is assumed to be 0.02. The parameter to be updated is the stiffness of each spring of the system. Three model parameters $\boldsymbol{\theta} = [\theta_1, \theta_2, \theta_3]^T$ are used to parameterize the model stiffness, defined as $k_i = k_0^i \theta_i$, and representing the product of nominal stiffness and model parameters. A Gaussian distribution with hyper mean $\mu_{\boldsymbol{\theta}} = [1,1,1]^T$ and hyper covariance matrix $\boldsymbol{\Sigma}_{\boldsymbol{\theta}} = diag(0.05^2, 0.05^2, 0.05^2)$ is used to simulate the measured dataset. The acceleration measurements are generated as the response for further updating purpose. Sensors are located on all three DOFs so that all the measurements can be considered for parameter estimation. Each dataset $D_i$ is contaminated with 2% Gaussian white noise in order to add measurement error.

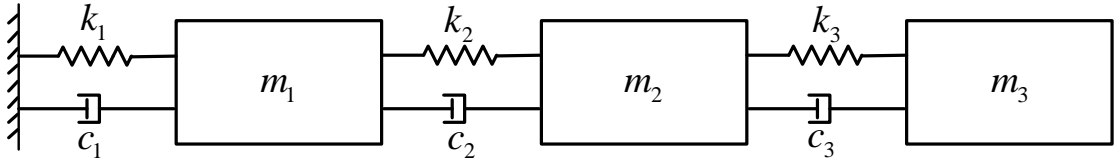

**Fig. 5** 3-DOF mass spring system

The two-stage algorithm introduced in Section 3.2 is used to compute the posterior distribution of the hyper parameters. The number of the samples of the hyper parameters is specified as 5000. Four cases with different number of datasets ranging from $N_D = 10$ to $N_D = 500$ are explored to compute the posterior distribution of the hyper parameters. In particular, the means along with the standard deviations of the hyper parameters are reported in **Table 3**. It can be observed that the mean values of the hyper mean and hyper standard deviation of the model parameters and prediction errors obtained from different data sets are close to the nominal values. However, when given a smaller data set ($N_D = 10$), the mean of the posterior sample deviates from the nominal value of the hyper parameter. Therefore, the hyper parameters have large uncertainty. With the increase of the number of data sets, the posterior sample mean is closer to the nominal value of the hyper parameter, and the uncertainty of the hyper parameter decreases. The above results can prove that with the increase of the data set, the uncertainty row of the hyper parameter decreases, and when the data set is large enough, the mean value of the hyper parameter will be closer to its true value. In addition, the PDF of the hyper parameter posterior sample corresponding to the model parameters of different datasets is shown in **Fig. 6**. It can be observed that when a small dataset ($N_D = 10$) is used, the mean of the posterior sample deviates significantly from the nominal value of the hyper parameter, and the uncertainty associated with the hyper parameter is relatively large. However, the nominal value of the hyper parameter still lies within the range of this uncertainty. As the size of the dataset increases, the uncertainty of the hyper parameters decreases, and the sample mean of most parameters approaches their nominal values. This trend is consistent with the results presented in **Table 3**. An increase in the number of datasets leads to a narrower range for the probability density function, meaning the distribution of the samples becomes denser, thereby reducing uncertainty. **Fig. 7** further illustrates the posterior distributions of the hyper parameters and prediction error variance for case $N_D = 500$. These posterior samples will subsequently be used for reliability updating purposes.

With the samples of the hyper parameters from **Fig. 7**, the distribution of the model parameters could be obtained. Herein, two cases from the HBM framework are considered, where the first case only considers the mean of the hyper parameters with neglecting their uncertainties while the second case accounts for all the samples of the hyper parameters. These two cases are also compared with the result from CBM approach where all datasets are combined for parameter inference. **Fig. 8** shows the parameter distribution using the HBM and CBM approaches with HBM considering two cases mentioned above. It is seen that the results from the two cases in the HBM framework are quite similar, primarily because the number of datasets is large, and the uncertainty of the hyper parameters does not play a significant role. In contrast, the CBM approach produces a highly peaked distribution, as it underestimates parameter uncertainties. This narrow distribution is unreasonable when accounting for parameter variability, as it fails to properly reflect the inherent uncertainty in the parameters.

**Table 3** Estimate the mean value of the hyper parameters using the HBM method

| | | $\theta_1$ | | $\theta_2$ | | $\theta_3$ | | $\sigma$ | |
|---|---|---|---|---|---|---|---|---|---|
| | | $\mu_{\theta_1}$ | $\sigma_{\theta_1}$ | $\mu_{\theta_2}$ | $\sigma_{\theta_2}$ | $\mu_{\theta_3}$ | $\sigma_{\theta_3}$ | $\mu_\sigma$ | $\sigma_\sigma$ |
| $N_D = 10$ | Mean value | 1.0179 | 0.0606 | 1.0354 | 0.0542 | 1.0232 | 0.0584 | 0.0202 | 0.0002 |
| | Standard deviation | 0.0205 | 0.0122 | 0.0216 | 0.0122 | 0.0198 | 0.0142 | $1.2570 \times 10^{-4}$ | $1.3321 \times 10^{-4}$ |
| $N_D = 50$ | Mean value | 0.9896 | 0.0410 | 0.9919 | 0.0432 | 0.9923 | 0.0415 | 0.0199 | 0.0002 |
| | Standard deviation | 0.0058 | 0.0043 | 0.0061 | 0.0050 | 0.0058 | 0.0042 | $5.1974 \times 10^{-5}$ | $9.4092 \times 10^{-5}$ |
| $N_D = 200$ | Mean value | 0.9935 | 0.0451 | 0.9933 | 0.0454 | 0.9933 | 0.0453 | 0.0200 | 0.0001 |
| | Standard deviation | 0.0032 | 0.0023 | 0.0032 | 0.0023 | 0.0031 | 0.0023 | $2.7965 \times 10^{-5}$ | $5.6624 \times 10^{-5}$ |
| $N_D = 500$ | Mean value | 1.0032 | 0.0525 | 1.0035 | 0.0524 | 1.0035 | 0.0523 | 0.0200 | 0.0001 |
| | Standard deviation | 0.0024 | 0.0017 | 0.0023 | 0.0015 | 0.0023 | 0.0016 | $1.6281 \times 10^{-5}$ | $3.2013 \times 10^{-5}$ |

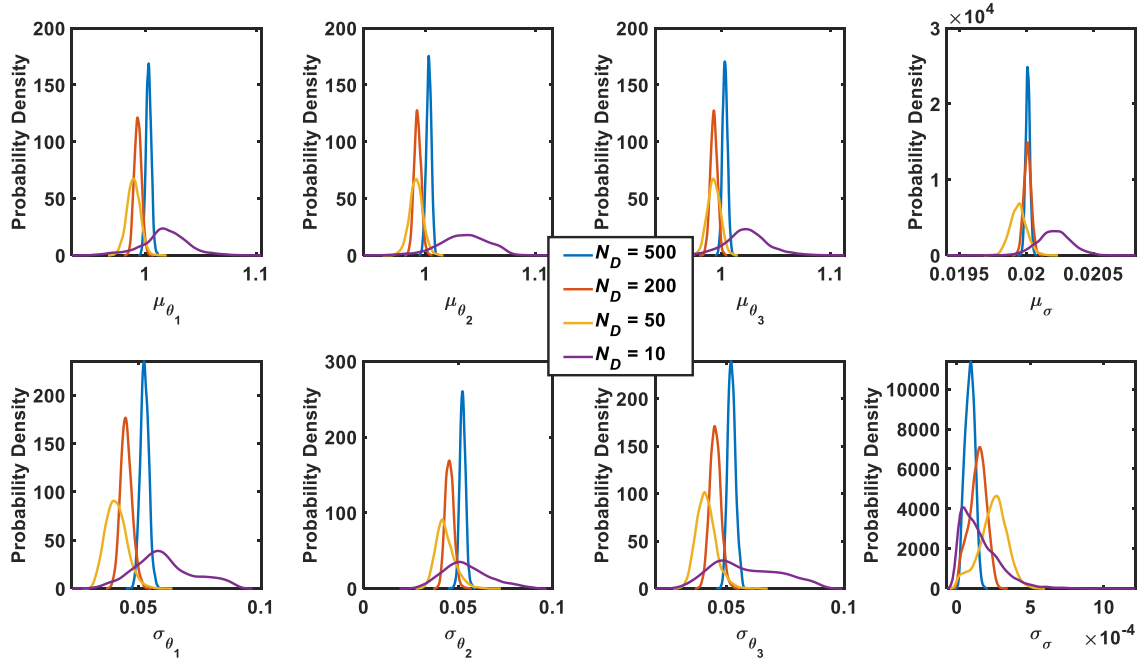

**Fig. 6** Distributions of the hyper parameters corresponding to $N_D = 10, 50, 200$ and $500$

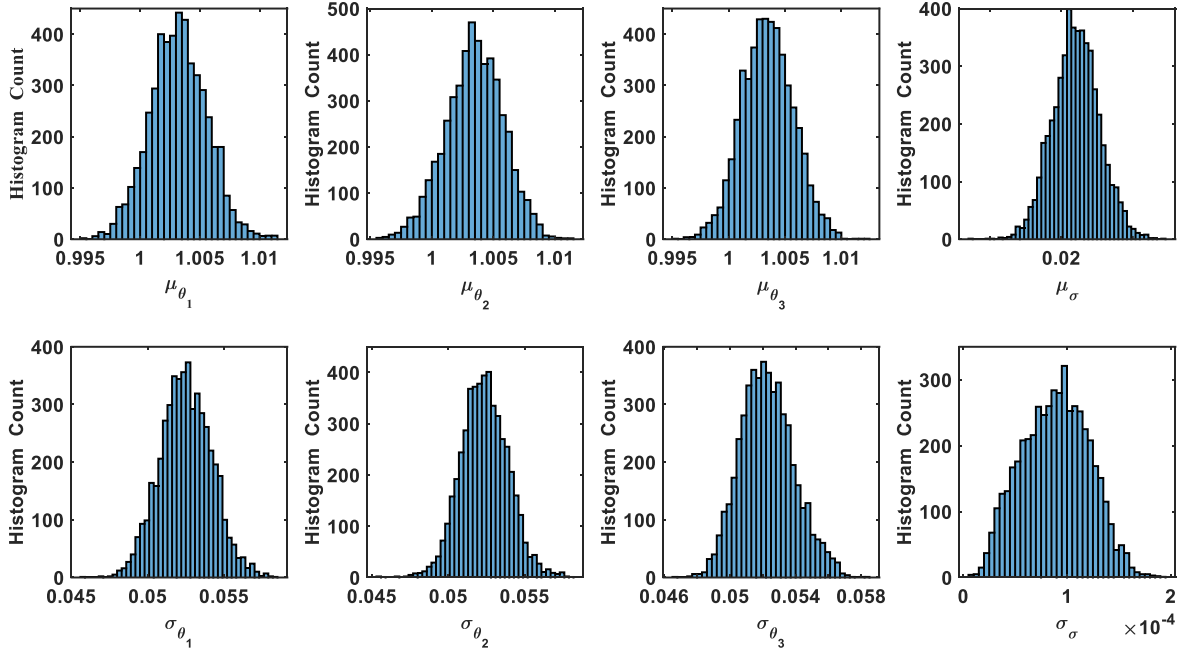

**Fig. 7** Posterior distribution of the hyper parameters using HBM method ($N_D = 500$)

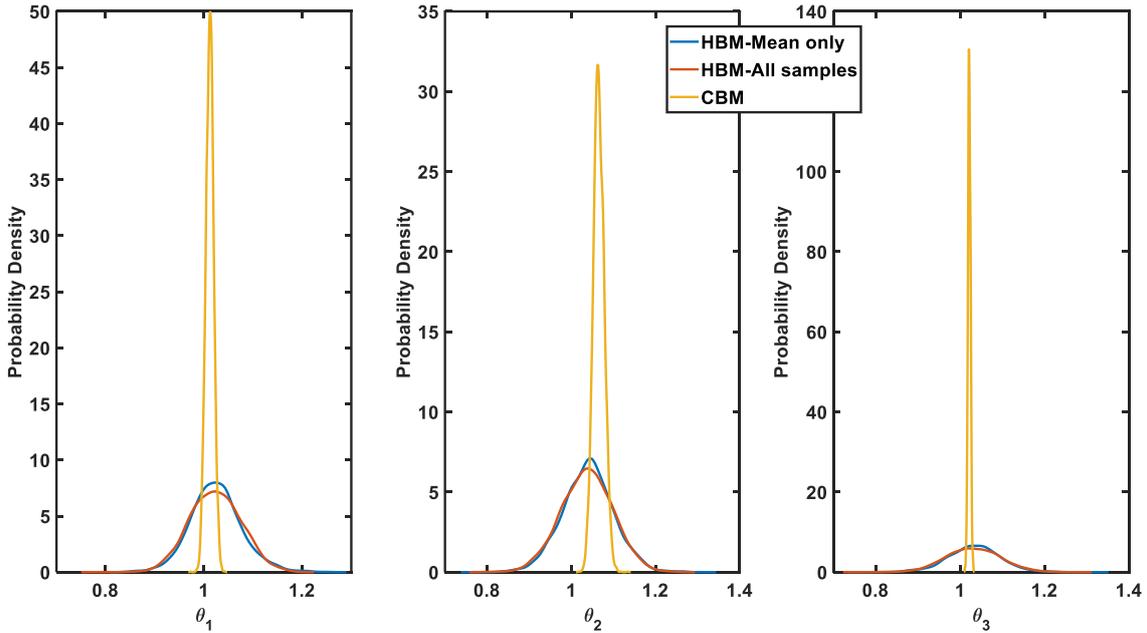

**Fig. 8** Model parameter distribution

In addition, the probability that the maximum displacement exceeds a predefined threshold is calculated. For this purpose, the limit state function is defined as $G(\boldsymbol{\theta}, \boldsymbol{\varphi}) = d_i(\boldsymbol{\theta}, \boldsymbol{\varphi}) - d_0$ under the constraint that the displacement of the system does not exceed $d_0$, where $d_i(\boldsymbol{\theta}, \boldsymbol{\varphi})$ is the displacement of the *i*-th DOF, and $d_0$ is the threshold. The uncertainty of the input is also considered which is generated as a white noise sequence with 1000 parameters having a mean of zero and a standard deviation of one. **Fig. 9** presents three sets of results: two from the HBM framework and one from the CBM approach. In the HBM framework, the results are similar because a large number of datasets are used, leading to relatively peaked distributions of hyper parameters. In contrast, the CBM approach shows markedly different results, giving predictions that are essentially one to many orders of magnitude different compared to the first two. The reason for this difference is that the CBM method seriously underestimates the uncertainty. These differences are further illustrated in **Fig. 10**, which shows the predicted probability density function of the maximum displacement of the system. The CBM-based PDF is the narrowest, clearly indicating that CBM significantly underestimates uncertainty compared to the HBM framework. On the other hand, the PDF based on the HBM hyperparameter sample mean is consistent with the prediction obtained using all hyperparameter samples as sufficient number of datasets used in this case.

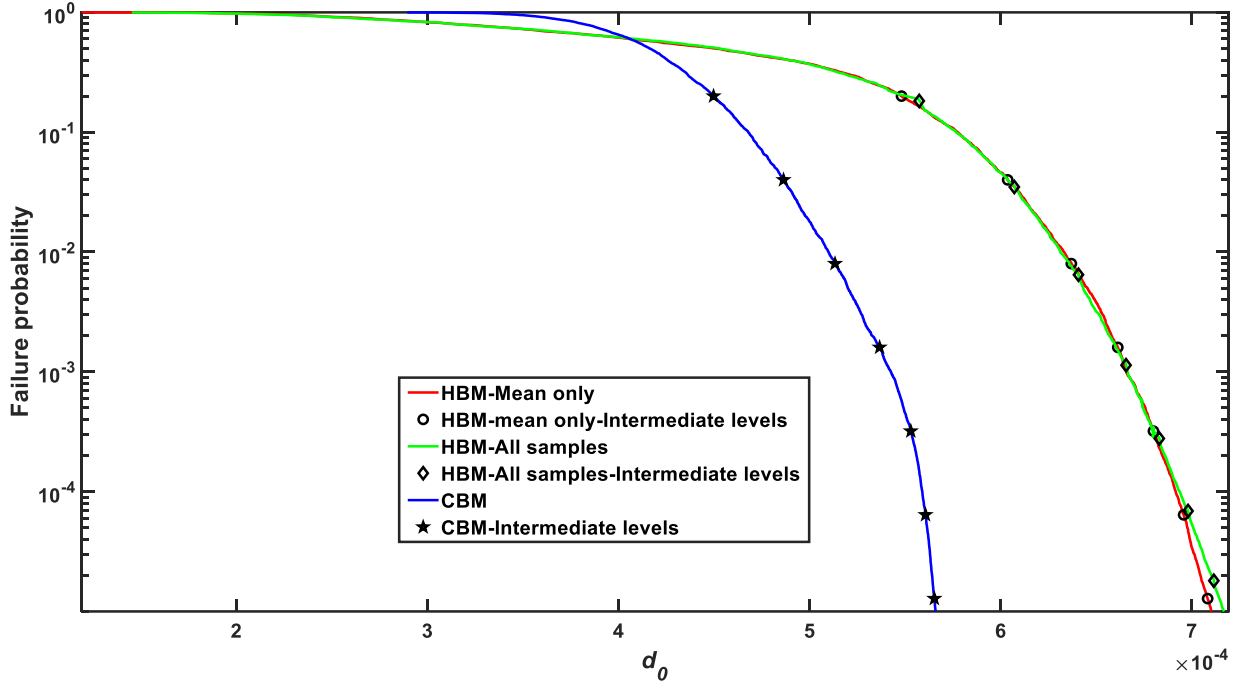

**Fig. 9** The failure probability of the maximum displacement

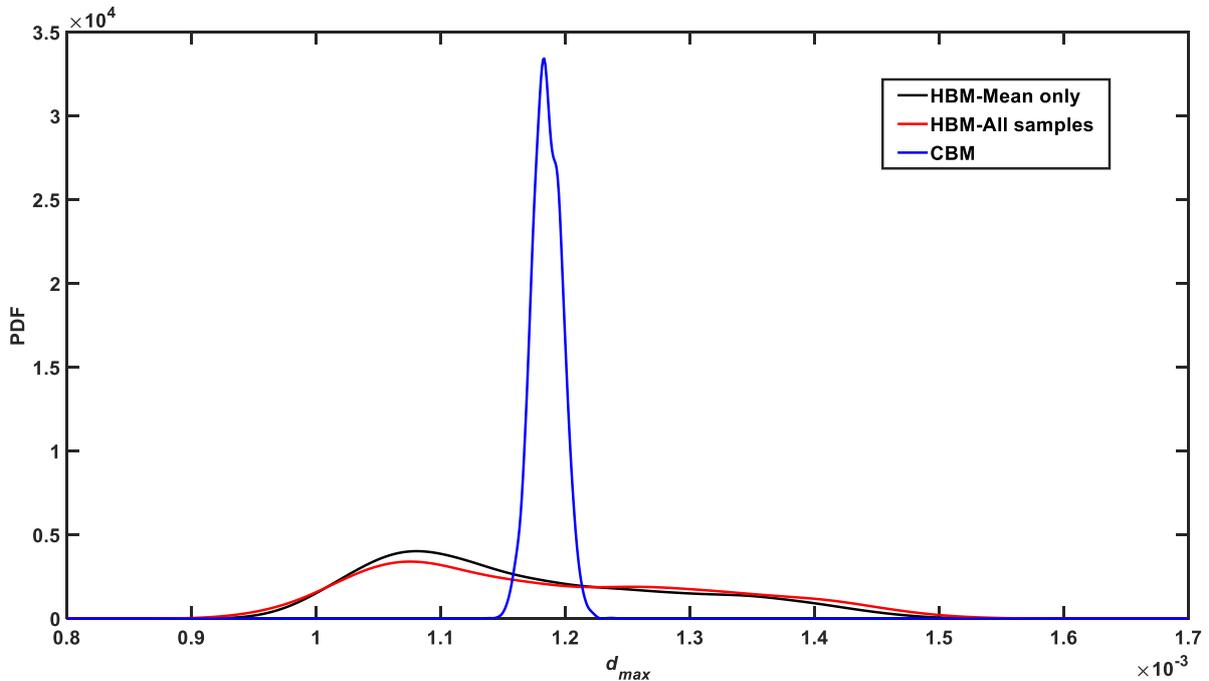

**Fig. 10** Posterior prediction distribution of maximum displacement

## 5. Conclusion

This study developed a HBM framework for uncertainty quantification and reliability updating using data, with applications to a linear mathematical model and a dynamical structural model. For the linear model, an analytical solution was derived for the hyper parameters and reliability index, demonstrating the efficiency and precision of HBM in simpler cases. For the dynamical structural model, where analytical solutions are not feasible, a two-stage sampling approach was introduced to ensure the accuracy of the framework. The results highlight a critical difference between HBM and CBM in how uncertainty influences reliability predictions. While CBM underestimates uncertainty as more datasets are incorporated, often leading to overly optimistic and unrealistic reliability predictions, HBM maintains a realistic representation of uncertainty, resulting in reliability predictions that are robust and reflective of the true variability in the system. This work underscores the advantages of HBM for addressing challenges in reliability assessment and highlights its potential for further applications. Future work will focus on extending this framework to incorporate datasets collected over periodic intervals, allowing for continuous reliability updates. Additionally, efforts will be made to integrate damage detection into the framework, enabling more comprehensive monitoring and assessment of structural health. These advancements aim to enhance the capability of the HBM framework to address real-world engineering challenges, particularly in systems subject to evolving operational and environmental conditions.

**Acknowledgment:** This work is supported by the National Natural Science Foundation of China (Grant 52305255), the Natural Science Foundation of Hebei Province (Grant E2023202066) and the S&T Program Hebei (24464401D).